# Accelerate & Actualize: Can 2D Materials Bridge the Gap Between Neuromorphic Hardware and the Human Brain?

Xiwen Liu[1]*, Keshava Katti[1]* Deep Jariwala[1]**.

**Affiliations:**

[1]Department of Electrical and Systems Engineering, University of Pennsylvania, Philadelphia, PA, USA.

*Co-first author. Email: xiwenliu@seas.upenn.edu, kattikes@seas.upenn.edu.

**Corresponding author. Email: dmj@seas.upenn.edu.

**Progress and Potential:**
- Two-dimensional (2D) materials present an exciting opportunity for devices and systems beyond the von Neumann computing architecture paradigm due to their diversity of electronic structure, physical properties, and atomically-thin, van der Waals structures that enable ease of integration with conventional electronic materials and silicon-based hardware.
- All major classes of non-volatile memory (NVM) devices have been demonstrated using 2D materials, including their operation as synaptic devices for applications in neuromorphic computing hardware.
- Their atomically-thin structure, superior physical properties, i.e., mechanical strength, electrical and thermal conductivity, as well as gate-tunable electronic properties provide performance advantages and novel functionality in NVM devices and systems. However, device performance and variability as compared to incumbent materials and technology remain major concerns for real applications.
- Ultimately, the progress of 2D materials as a novel class of electronic materials and specifically their application in the area of neuromorphic electronics will depend on their scalable synthesis in thin-film form with desired crystal quality, defect density, and phase purity.

**Graphical Abstract:**

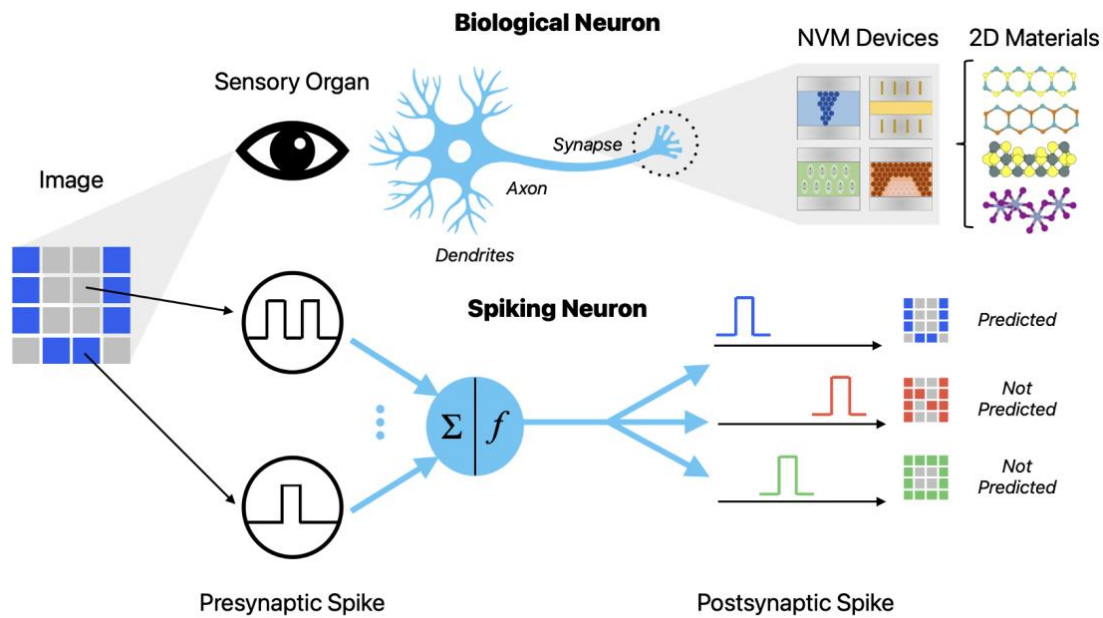

**One-Sentence Summary:** Integration of 2D materials into emerging memory devices leads to low-power, gate-tunable, analog computing architectures that can both accelerate existing machine learning algorithms and actualize neurobiological functions for brain-like learning.

# 1. Introduction

Neuromorphic computing broadly represents the use of non-von Neumann architectures to emulate learning exhibited by the human brain. The term "von Neumann architecture" represents any stored-program computer in which the fetch instruction and data operation may not occur simultaneously since they share a common bus, leading to the "von Neumann bottleneck" that describes the energy- and time-intensive transfer of data between separate memory and compute blocks. Such a bottleneck restricts the ability of computing systems to execute data intensive tasks whose demands are only growing with the advent of modern machine learning models. Furthermore, a recent report shows that highly complex neural networks operating in the "over-parameterized regime" do not overfit to spurious trends in training data but rather exhibit better generalization to unseen data than their less complex counterparts [1], motivating exponential growth each year in the number of model parameters since 2015 and the size of training data sets since 1988 [2, 3]. In particular, the past decade has witnessed models from ResNet-50 ($> 10^7$ model parameters) to the Generative Pre-trained Transformer 3 (GPT-3) ($> 10^{11}$ model parameters) and datasets from ImageNet ($\sim 10^6$ images) to JFT-3B ($> 10^9$ images). By overcoming this bottleneck in electronic communication, clocking, thermal management, and power delivery [2], neuromorphic systems bring the promise of scalable hardware that can keep pace with the exponential growth of deep neural networks, leading us to define the first major thrust of neuromorphic computing: "acceleration." Those neuromorphic systems concerned with acceleration are built for heightened speed and energy efficiency in *existing* machine learning models and tend to have a relatively immediate impact. One common example would be crossbar arrays for vector-matrix multiplication (VMM) in the forward pass of deep neural networks. In contrast, we define the second major thrust of neuromorphic computing to be "actualization," which is the realization of human neurobiological functions in non-von Neumann architectures. The second thrust will have a more delayed impact than the first but represents the hardware implementation for next-generation machine learning models, with headway being made in the space of spiking neural networks (SNNs), Hebbian learning, and Hodgkin-Huxley neuron models.

Both acceleration and actualization approaches to neuromorphic computing require development of specialized hardware. Memory technologies that support reading and writing can be historically categorized as either random-access memory (RAM) or flash memory. RAM is volatile, meaning it requires a power supply to refresh its memory states every few milliseconds, leading to undesirably large power consumption for neuromorphic applications [4]. While silicon-based flash memory represents the dominant non-volatile memory (NVM) device in the electronics industry, it also suffers from deficiencies with respect to neuromorphic systems, including slow programming speed, relatively high operating voltage, and poor electrical endurance [5]. Furthermore, both RAM and flash memory are based on charge storage, which can use either a capacitor, cross-coupled inverter, or floating-gate transistor, and are currently struggling with scalability below 10 nm lateral dimensions. Both memories are also separate from the compute block, meaning they do not offer a solution to the von Neumann bottleneck. A new class of memory, often referred to as "emerging memory devices," have been gaining much traction over the past decade due to their combination of fast switching speed in static RAM (SRAM), large storage density in dynamic RAM (DRAM), and non-volatility in flash memory, as well as their multi-bit/analog in-memory computing abilities.

Emerging NVM devices can be categorized into four different groups: resistive, phase-change, ferroelectric, and ferromagnetic. Resistive devices exhibit resistive switching, which is a cyclical change of device resistivity via formation/rupturing of a conductive filament across distinct stable levels as induced by an electrical bias [6] (Fig. 1ai). For a two-state resistive device, the low-resistance state (LRS) is achieved applying a positive voltage $V > V_{set}$, while the high-resistance state (HRS) is achieved by applying a negative voltage $V < V_{reset}$ (Fig. 1bi). Phase-change memory (PCM) is based on electrical contrast between highly resistive amorphous states and highly conductive crystalline states in phase-change materials [7] (Fig. 1aii). In order to change the state of a phase-change device, high-power heat pulses are required to melt/quench the material between crystalline and amorphous states (Fig. 1bii). Ferroelectric devices leverage the change in current or resistance due to the induced polarization of a ferroelectric material with respect to a voltage bias [8] (Fig. 1aiii). A two-state ferroelectric device requires a voltage $V > +V_{coercive}$ to induce a remnant polarization $+P_r$ in the ferroelectric material and a voltage $V < -V_{coercive}$ to induce a remnant polarization $-P_r$ in the ferroelectric material (Fig. 1biii). Ferromagnetic devices contain two ferromagnetic layers, a reference layer that is pinned to a fixed polarization and a free layer can take on a programmed polarization [8] (Fig. 1aiv). These two layers can either be parallel or anti-parallel, which induces a LRS or HRS, respectively (Fig. 1biv).

This review focuses on the promising subset of emerging NVM devices for which two-dimensional (2D) crystals form a critical component, as well as those devices that are entirely made from 2D crystals and their van der Waals heterostructures. Following the advent of stable monocrystalline graphitic films at the thickness of only a few atoms [9], the search for more stable 2D materials has yielded a new generation of emerging memory devices that offer even higher switching speed ($< 10$ ns) and lower power loss ($\sim 10$ pJ) while maintaining a low threshold voltage ($< 1$ V) and lengthy retention time ($> 10$ years) [5]. The additional knob of gate-tunability enabled by 2D materials-based active layers is another exciting advantage for neuromorphic applications, as it gives rise to a host of multi-state/analog memory devices. The goal of this perspective will be to connect how 2D materials engender better memory devices that in turn create architectures and algorithms for next-generation machine learning models. We will then hypothesize about the future direction of neuromorphic computing and illustrate the chasm that must be bridged between current machine learning hardware and the human brain.

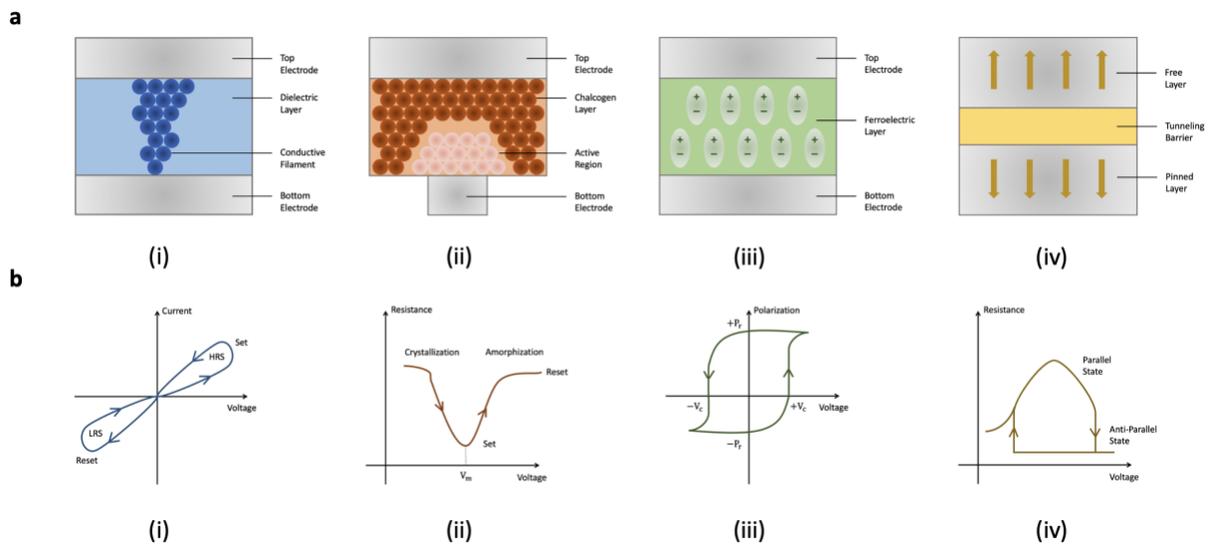

**Fig. 1. Four types of two-terminal emerging memory devices.** (**a**) Typical structures for (*i*) resistive device in low-resistance (1-bit) state with formed conductive filament, (*ii*) phase-change device in high-resistance (0-bit) amorphous state, (*iii*) FTJ with applied $V < -V_c$ in $-P_r$ (0-bit) state, and (*iv*) MTJ in high-resistance (0-bit) anti-parallel state. (**b**) Memory state change plots showing (*i*) current-voltage relationship following anti-clockwise loop that toggles between high-resistance and low-resistance states in resistive device, (*ii*) resistance-voltage relationship that outlines transition from low-resistance crystalline state and high-resistance amorphous state in phase-change device, (*iii*) anti-clockwise polarization-voltage hysterisis loop that toggles between $+P_r$ and $-P_r$ polarization states in FTJ, and (*iv*) clockwise resistance-voltage hysterisis behavior that toggles between low-resistance parallel state and high-resistance anti-parallel state in MTJ.

## 2. Past

The human brain is known for its high level of energy efficiency and low latency in processing complex information. It is able to perform better than any computer, especially for tasks that require processing large amounts of complex information, such as perception. These benefits are largely due to the brain's dense network of neurons that transmit signals efficiently through synapses. Some researchers saw the potential for electronic technology to model human brain system. For example, Hodgkin and Huxley modeled the currents carried by sodium and potassium ions through nerve axons in a squid as equivalent electrical circuits using a set of exponential equations in 1952 [10, 11]. A decade later, FitzHugh and Nagumo modeled threshold phenomena in the nerve membrane via analog electrical component [12, 13]. These electronic models allowed researchers to study and compare neural and electrical systems but were not designed for practical use [11]. Neuromorphic engineering is a field that seeks to replicate von Neumann architecture in silicon electronics. The term was coined by Carver Mead in the late 1980s to describe systems that use analog and digital circuits to mimic the neurobiological elements found in the nervous system [14].

Modern neuromorphic hardware platforms are based on a variety of technologies, including complementary metal-oxide-semiconductor (CMOS) field-effect transistors (FETs), memristors, and flash memory, which are essentially floating-gate transistors, among others. One key feature of neuromorphic hardware platforms is their ability to perform computations in a "spiking" manner, meaning they process data using discrete pulses of activity, similar to how neurons in the brain transmit information. These spikes allow neuromorphic hardware to perform computations more efficiently than traditional digital circuits, which process data using a continuous stream of bits. Research in this field at the architecture level is quite advanced but uses the same basic building blocks based in silicon CMOS. For instance, IBM's TrueNorth neuromorphic chip developed in 2014 has 4,096 neurosynaptic cores, each of which has 256 neurons and 256 synapses to communicate with other neurons, and is significantly more energy-efficient on tasks such as face recognition, collision avoidance, and eye detection than a conventional von Neumann processor, requiring as little power as 65 mW [15]. The major difference in power consumption between TrueNorth and a conventional von Neumann processor is due to TrueNorth's asynchronous, event-driven processing. In this type of processing, only neurons that are actively being used are updated at each time step, rather than all neurons being updated constantly. In 2019, the University of Manchester's Advanced Processor Technologies (APT) group developed the world's fastest supercomputer, called SpiNNaker, made up entirely of neuromorphic cores [16]. Akida, developed by the company Brainchip, is the first commercial neuromorphic hardware, available since 2021 [17], while Intel's Loihi is a state-of-the-art neuromorphic processor with 128 cores, each containing 1,024 neurons, developed by Intel to model SNNs in silicon CMOS technology [18]. Fabricated using Intel's 14 nm CMOS process and measuring 60 mm$^2$, Loihi represents a significant advancement in the field of neuromorphic computing and is capable of processing large amounts of data in real-time. Operation of the Loihi chip can be divided into three main components: neurons, synapses, and interconnections between them. First, neurons receive input and generate output pulses, or "spikes," based on this input using an activation function that can be programmed to perform various tasks such as thresholding, filtering, or non-linear processing. Second, synapses are used to store and transmit information between neurons and can be programmed to have different levels of "weight," determining the strength of the connection between neurons. Third, interconnections are used to transmit spikes between neurons, as well as to store and retrieve weights in synapses.

In addition to being implemented in CMOS technology, neuromorphic systems have also been created using NVM technologies. These systems are attractive because the underlying physics of NVM devices can be directly linked to certain properties of biological neurons and synapses, allowing a single NVM device to function as a neuron or synapse with varying levels of activity [8, 19, 20]. Further, NVM devices are typically two- or three-terminal and end up consuming much less power when mimicking the same biological property of a neuron in comparison to CMOS technology. NVM technologies can also be arranged in dense crossbar arrays of synapses with neurons at the periphery, a key feature of the biological brain that distinguishes it from traditional computing systems with separate compute and storage units. Crossbar architectures have the potential to store a large amount of data with a small physical footprint, making them useful for building large-scale neuromorphic systems that require complex neurotrophic algorithms [19]. In addition, crossbar architectures are known for their ability to perform VMM very quickly using Kirchhoff's Current Law (KCL) [8]. By applying the appropriate voltages to the NVM devices in a crossbar architecture, VMM can be performed in a single time step, resulting in superior computing throughput. In particular, the input matrices are first encoded into a series of voltages which are then applied to the word lines in the crossbar. The resulting currents flowing through the memristors are then measured at bit lines and used to calculate the elements of the output matrix. VMM is particularly useful for deep learning algorithms, which place tremendous demands upon von Neumann digital systems. Crossbar architectures can also directly interface with analog signals from sensors, making them suitable for various applications, including edge computing and robotics.

Numerical simulations have shown that crossbar architectures can achieve speed and energy efficiency that are orders of magnitude higher than traditional CMOS hardware platforms [21, 22]. Crossbar architectures have also been demonstrated in physical hardware. For instance, recent report shows that an integrated array of 1,024 NVM devices can be used for gray-scale face classification, with energy consumption significantly lower (1,000x) than that of the Intel Xeon Phi processor and comparable accuracy to a central processing unit (CPU) [23]. Another report demonstrates that memristor crossbars made of $HfO_2$ memristors on top of MOS transistors can perform analog VMM with array sizes of up to $128 \times 64$ cells for signal processing, image compression, and convolutional filtering applications [24]. These crossbars have an output precision of 5-8 bits due to high device yield of 99.8% and multilevel, stable states. Another instance is a mixed hardware-software neural network that incorporates up to 204,900 phase-change memory synapses [25]. This network is able to achieve generalization accuracies on commonly used machine learning test datasets that are equivalent to those of software-based training, with throughput being two orders of magnitude higher than current graphical processing units (GPUs). Lastly, another study shows the use of a diffusive memristor to create an artificial neuron with stochastic leaky integrate-and-fire (LIF) dynamics and tunable integration time, exhibiting experimental evidence of unsupervised synaptic weight updating and pattern classification [26].

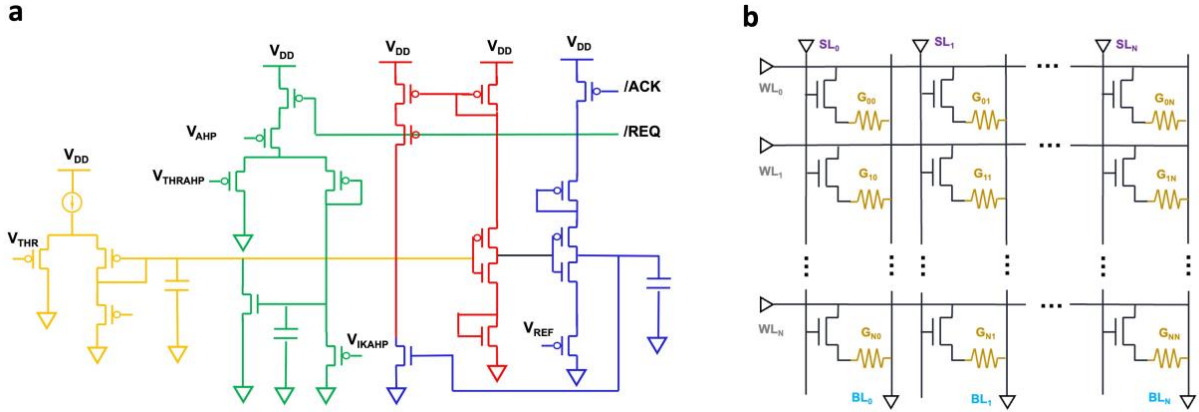

**Fig. 2. Side-by-side comparison of CMOS neuron circuit and neuromophic crossbar array with NVM.** (**a**) Schematic of integrate-and-fire neuron circuit on CMOS technology. The sub-circuit in yellow models the neuron's leak conductance; the sub-circuit in green implements the spike event generation process; the sub-circuit in red is used to reset the neuron for a refractory period; the sub-circuit in blue implements integration over the spikes [27]. (**b**) Schematic of a neuromophic crossbar array with NVM. The SL denotes search line, WL denotes word line, BL denotes bit line, and G denotes the conductance of the NVM devices, which can be dynamically programmed by electrical signals.

## 3. Present

The performance of memory cells is a crucial factor for in-memory computing systems. In this section, we will review the most recent developments in 2D memories for in-memory neuromorphic computing technology, including device performance, design considerations, and potential applications. 2D memories can be classified into four main categories based on their working mechanisms: resistive devices, phase-change devices, ferroelectric devices, and ferromagnetic devices. There are several key factors that are used to evaluate the performance of 2D memory technologies, including the speed at which data can be written or accessed (write/access time), the length of time that data can be stored (retention time), the amount of power energy consumed during operation (power/energy consumption), the number of times that the memory can be cycled before it begins to degrade (cycling endurance), the ratio of read to write operations that the memory is capable of performing (ON/OFF ratio), the ability to store multiple states of data (multiple states), and the ability to expand or scale the memory capacity as needed (scalability).

*3.1. Resistive Devices*

Resistive memory, also called Resistive RAM (ReRAM) or resistive switching memory, is a type of non-volatile memory that uses resistance to store and retrieve data. Advanced and commercial-scale versions of ReRAM devices consist of a thin-film resistive material, such as $TiO_x$, $TaO_x$, or $HfO_x$, that can be switched between a LRS and HRS by applying a voltage. The HRS corresponds to a 0-bit, while the LRS corresponds to a 1-bit, allowing for data storage and retrieval [28]. ReRAM has several advantages over traditional NVM technologies, like flash memory, including higher density, faster read/write speeds, higher endurance, and lower power consumption. However, there are challenges that have hindered the use of ReRAM in large-scale applications, including non-linearity and asymmetry in switching, device-to-device variation, and low ON/OFF ratio. Additionally, the high write energy required for ReRAM has been a limiting factor.

Recent research has examined the potential for resistive switching in 2D materials to address challenges in the development of ReRAM technology [29-31]. Some studies have found that single-layer $MoS_2$ sandwiched between metal electrodes can exhibit stable non-volatile resistance switching at operating voltages as low as 0.3 V [32]. However, the endurance of these devices is $< 10^3$ cycles, and the mechanism behind the resistive switching is not fully understood. To improve the reliability of 2D materials-based ReRAM, another report uses van der Waals heterostructures to stack resistive layers and electrodes, resulting in devices with endurance of up to $10^7$ cycles and the ability to operate at high temperatures ($> 600$ K) [33]. Furthermore, 3D monolithic integration of a 1T-1R memory cell using a 2D $WSe_2$ transistor and a solution-processed $WSe_2$ ReRAM was demonstrated, which offers potential for 3D embedded memories in emerging computing systems [34]. Variability in 2D materials has also been a challenge in the creation of high-density electronic circuits for complex applications, but a recent report presents a method for fabricating high-density ReRAM crossbar arrays using h-BN as the resistive switching material, which achieves a high yield of 98% and low cycle-to-cycle (1.53%) and device-to-device variability (5.74%) [35]. Another group went on to develop a 2D ReRAM-based crossbar architecture using polycrystalline $HfSe_2$ and a metal-assisted van der Waals transfer technique, which exhibited low switching voltage (0.6 V) and energy (0.82 pJ). The reported crossbar array was able to emulate synaptic weight plasticity with a high recognition accuracy of 93.34%, as well as boast a high-power efficiency of more than 8 trillion operations per second per watt [36].

*3.2. Phase-Change Devices*

In conventional PCM technology [7], phase-change materials, such as $Ge_2Sb_2Te_5$ (GST), can be rapidly and reversibly switched between a high-resistance amorphous phase and a low-resistance crystalline phase by applying vertical voltages that cause Joule heating. Although PCM has the ability to operate at high speeds and has a high endurance compared to flash memory, it requires a relatively high write current or voltage for programming due to its switching mechanism relying on thermally-induced phase transformation. Fortunately, the use of 2D materials in PCMs has opened up new opportunities for research into novel phase change mechanisms and has the potential to bring significant performance improvements [30]. One benefit of using 2D materials in PCMs is their high electrical conductivity and significantly anisotropic thermal conductivity, which can improve the speed and energy efficiency of phase change processes. A report from 2015 shows that by inserting a layer of graphene between GST and the bottom electrode heater, PCM devices show a 40% reduction in reset current with minimal increase in electrical resistance compared to control devices without the graphene layer [37]. The improvement in energy efficiency is thought to be due to the graphene layer acting as an additional thermal resistance at the interface, helping to keep the heat generated during the phase change process confined within the active volume of the PCM.

In addition to the performance enhancements that can be achieved by incorporating 2D materials into traditional GST PCM devices, the exploration and manipulation of new phases of these materials have also opened the door to new avenues of research in novel phase change mechanisms [38]. A 2019 report shows that in vertical devices that use 2D materials of $MoTe_2$ and $Mo_{1-x}W_xTe_2$, the application of an electric field across the film thickness direction can cause a transition in the MX$_2$ material from the 2H phase to a distorted, transient structure known as the 2H$_d$ phase [39]. In these devices, resistive switching between the LRS and HRS is observed to occur within 10 ns and is reproducible. In addition, an Al$_2$O$_3$/MoTe$_2$ stack PCM has been demonstrated with ON/OFF current ratios of $10^6$ and programming currents lower than 1 mA. A recent report also developed a new type of PCM by using a layered crystalline 2D In$_2$Se$_3$ film on a graphene bottom electrode and leveraging the reversible phase changes between the β and γ phases for SET/RESET programming [40]. In these devices, it was observed that resistive switching between high and low resistive states occurred reproducibly within 50 ns and was stable up to 100 cycles. The ON/OFF current ratio was measured to be 50, and the programming current was higher than 100 μA. In addition to being used in vertical PCM devices, 2D phase change materials have also been implemented in in-plane devices. Another report shows that it is possible to reversibly modulate MoS$_2$ films by applying an electric field to control the migration of Li+ ions, leading to local 2H–1T' phase transitions, in which an increase or decrease in the local concentration of Li+ ions causes the material to transition between the 2H (semiconductor) and 1T' (metal) phases [41]. The device also demonstrated a high ON/OFF current ratio of 100, as well as strong endurance and retention up to 40,000 cycles. Additionally, the conductance of the device can be adjusted in incremental steps by applying short voltage pulses for more than 200 states, enabling the device to mimic the behavior of synaptic potentiation and depression.

*3.3. Ferroelectric Devices*

The three main types of ferroelectric memory include ferroelectric RAM (FeRAM), FE-FETs, and ferroelectric tunnel junctions (FTJs). FeRAM is composed of a one transistor and one capacitor (1T-1C) structure. The significant downside of FeRAM is its destructive read operation, meaning its state needs to be rewritten after a read operation is executed, similar to the constant refreshing in standard RAM. In contrast, FE-FETs and FTJs use non-destructive read operations. While FTJs leverage the direction of polarization to modulate device resistance, FE-FETs combine the 1T-1C structure of FeRAM into a single device by integrating a ferroelectric film into the gate stack of a MOSFET.

The first kind of neuromorphic architectures enabled by 2D materials are those leveraging either FE-FETs or FTJs with a 2D ferroelectric film. After ferroelectricity in CuInP$_2$S$_6$ (CIPS) of ~4 nm thickness was exhibited [42], a group of researchers constructed an FTJ with a Au/CIPS/Ti structure [43]. Their results suggest that 2D ferroelectric materials can avoid poor cycle-to-cycle and cell-to-cell variations of conductive filaments seen in RRAM devices, supported by small V$_{set}$ and V$_{reset}$ variation ($\sigma/\mu$) of 5.3% and 9.1%, respectively. Each FTJ represents an artificial synapse with synaptic weight corresponding to the device conductivity and spiking action potential achieved by applying sequential positive and then negative voltage pulses of equal number. Another report found ferroelectricity in a < 6 nm SnS film, employed within a Pt/SnS/Pt FTJ [44]. This device acts as an artificial synapse with a similar strategy of generating action potentials using device conductivity, achieving a high conductance ratio ($G_{max}/G_{min}$) of 20.5 with 100 conductance states. The authors use NeuroSim+ to simulate an array of FTJs comprising a two-layer fully connected neural network [45]. Forward-propagation is executed on 16,000 training images from the Modified National Institute of Standards and Technology (MNIST) dataset, a collection of handwritten digits from zero to nine. NeuroSim+ does not simulate back-propagation in hardware but implements all gradient computations in software and then loads the updated weights into the crossbar array. This report shows a 92.1% training accuracy after 100 epochs, suggesting the proposed FTJ suitably quantizes neural network weights for forward-propagation.

The second kind of devices are FE-FETs and FTJs integrated with a non-ferroelectric 2D active layer. One report demonstrates an HZO/WS$_2$-based FE-FET, where WS$_2$ is the 2D channel material helping the device achieve a high ON/OFF ratio of ~$10^5$ and change in postsynaptic current of > 10 $\mu A$ [46]. A recent example of a 2D MoS$_2$ channel in an FE-FET shows an immense ON/OFF ratio of ~$10^6$ and normalized memory window of 0.3 V/nm with the ferroelectric AlScN [47], while another instance involves the organic ferroelectric polyvinylidene fluoride-trifluoroethylene (PVDF-TrFE), showing stable memory states up to $10^7$ cycles, an ultra-low energy consumption of 1 fJ, and more than $10^9$ synaptic operations corresponding to a 10-year device lifetime [48].

*3.4. Ferromagnetic Devices*

Spintronics capture a subset of emerging devices that leverage intrinsic angular momentum of electrons by treating them as small bar magnets with strong magnetic moments [49]. Spin-based ferromagnetic devices can be classified into two categories: MTJs or spin transistors. The functional layer of an MTJ is made up of two ferromagnetic electrodes and a thin tunneling barrier sandwiched between them. An applied voltage to the MTJ modulates the two magnetic layers between being parallel (LRS) and anti-parallel (HRS). MTJs represent the building block of spin-transfer-torque-MRAM (STT-MRAM). The second category of ferromagnetic devices are spin transistors. Varying the gate voltage in spin transistors alters spin precession angle, inducing a change in spin polarization of current on the channel.

The spintronics community has found many applications of 2D materials and corresponding heterostructures in ferromagnetic devices. For example, 2D materials-based functional layers could alleviate issues with control over the tunneling barrier, thermal stability, and robustness, key blockades preventing further improvement of MTJ performance and scaling [50]. In particular, a theoretical study showed that an MTJ with a graphene-based tunneling barrier could achieve a tunneling magnetoresistance (TMR) ratio of 100% [51]. Yet the highest TMR ratio exhibited experimentally in a graphene-based MTJ remains only 31% [52]. Other 2D materials have been used as tunneling barriers, including h-BN, MoS$_2$, and WS$_2$, with h-BN exhibiting the best TMR ratio of 6% [53]. The latter two TMDCs show potential but have thus far achieved no better than a TMR ratio of 0.7% [49]. On the other hand, 2D ferromagnetic materials have fared better upon integration within spintronic devices. A recent report exhibits an MTJ with a four-layer 2D magnetic CrI$_3$ tunneling barrier, graphene contact, and h-BN encapsulation layer, showing a 19,000% TMR ratio at 2 K [54]. Other ferromagnetic heterostructures include FGT/h-BN/$FGT$, where FGT stands for Fe$_3$GeTe$_2$, and VSe$_2$/MoS$_2$/VSe$_2$, with the former yielding an experimental TMR ratio of 160% at 4.2 K and the latter theoretically predicting a TMR ratio of 846% at 300 K [55, 56]. Compared to MTJs, there is limited work on spin transistors integrated with 2D materials, with the most notable instance being a double-gated spin tunnel FET (sTFET) with a graphene/CrI$_3$/graphene heterostructure active layer, achieving a large conductance ratio of about 400% [57].

As the use of 2D materials in spintronic devices is still in its infancy [49], application of such ferromagnetic memory devices remains limited in literature, but the promise of better scalability, higher TMR ratio, longer spin lifetime, improved spin injection, and longer spin diffusion length are just a few of the potential benefits [58]. Thus far, most spintronics-based neuromorphic systems use MTJs in a crossbar array for VMM. One report shows inference in an *ex situ* neural network pre-trained in MATLAB, where the authors design their own STT-based non-linear neuron with a soft-limiting activation function [59]. The reported architecture achieves an energy consumption of ~650 fJ for a single alphabet recognition task, but no inference accuracy is reported. Neither do these authors integrate the MTJs with 2D materials nor have there been any other prevalent reports that do so, suggesting that the benefits of spintronics with 2D materials for neuromorphic applications are yet to be realized.

## 4. Problems

The above sections describe major recent developments in 2D materials-based NVM memory devices for application in neuromorphic computing architectures. While significant progress has been achieved, there are several major technological challenges that largely reside at the materials level. Though demonstrations at the individual-device level remain impressive, any complex microelectronic system requires more devices on the order of several magnitudes [60]. To move from device to architecture design, such NVM and synaptic devices must be exhibited at more than the single-device level. But in large-scale synthesis, which has been shown for some 2D materials, including graphene, $MoS_2$, and $WS_2$ [61-63], the crystal quality and defect density remains far from the desired level of control. Even in elemental 2D materials, such as graphene, crystalline quality resembling epitaxial Si and III-V crystals has only recently been demonstrated [64]. For binary 2D materials, such as h-BN and metal chalcogenides, which form the basis of most ReRAM and PCM demonstrations discussed above, such quality control remains far from the desired level. In addition, 2D materials that intrinsically undergo crystalline-to-crystalline phase changes or exhibit ferroelectricity or ferromagnetism are all at least binary compounds (e.g., MoTe2, In2Se3, $CrI_3$) or even ternary/quaternary in many cases, such as FGT, CIPS, and so on. Experience with binary, ternary and quaternary III-V and II-VI semiconductors suggests that achieving the desired composition, crystalline quality, and defect control for such 2D analogs is going to be equally challenging if not more given the state of research in vapor phase synthesis of 2D chalcogenides via chemical vapor deposition (CVD) and molecular beam epitaxy (MBE) techniques, but the outlook remains very optimistic of eventual success.

For 2D materials-based ReRAM devices, the low-switching voltages in ultrathin 2D layers is advantageous. However, the low endurance ($< 10^6$–$10^7$ cycles) of all demonstrations to puts them at a major competitive disadvantage when compared to 3D oxide memristors that are now commercially available and lie in the desirable corner of Fig. 3a. For PCM devices, the large write currents continue to pose challenges, even when the active material involved is atomically-thin. However, a potential advantage of 2D chalcogenide phase change materials over conventional 3D chalcogenides, such as GST and alike materials, is that many of them undergo a crystalline-to-crystalline phase transition in the presence of electric fields instead of temperature that requires passing large currents. Such a phase transition also involves minimal movement of atoms (e.g., $Mo_{1-x}W_xTe_2$ 2H to $2H_d$ phase transition) yet provides a disproportionately high change in resistivity potentially, pushing the device performance to the desired corner of Fig. 3b. However, the endurance stability, scaling, and multi-state programmability of such devices have been scarcely investigated and therefore remain a potential area of research and development.

2D materials-based ferroelectric and ferromagnetic devices face a unique set of challenges with respect to viability in neuromorphic architectures. FeRAM, FE-FETs, and FTJs all boast fast read/write speeds ($< 1$–$5$ ns), lengthy retention time ($> 10$ years), strong endurance ($> 10^7$–$10^{14}$ cycles), and relatively low write energy ($< 5$–$30$ fJ) [60]. Yet scaling remains one of the central challenges that can lower-bound power consumption and upper-bound data storage density, key device properties that determine the efficiency of a neuromorphic system. 1 nm scale Zr-doped $HfO_2$ exhibits limited polarization, as well as a dismal ON/OFF ratio of 10, likely due to the depolarization field and surface-related effects [65]. Poor switching behavior, low remnant polarization, and complicated design of lattice-matched growth substrates during fabrication remain stumbling blocks that prevent the effective realization of thin-film ferroelectrics in neuromorphic devices. While using 2D materials for the active layer of a ferroelectric device may improve scalability, it does not help downsize traditional bulk ferroelectrics, which can be up to hundreds of nanometers in thickness. Additionally, Fig. 3c shows most 2D materials-based ferroelectric devices exhibit an endurance of $< 10^5$, which is insufficient for the high-operation demands of neural networks. Therefore, most 2D materials-based ferroelectric NVM devices have stayed far from the desired corner though there are reasons to remain optimistic as detailed in the next section.

Exemplified by the limited number of reports showing neuromorphic architectures using 2D materials-based spintronic devices, there also exist a number of key limitations reducing the efficacy of ferromagnetic synaptic devices. As shown in Fig. 3d, the most notable of such limitations remains the very low operating temperature and permanence of 2D magnetic materials. While the TMR ratio of 19,000% seen in the $CrI_3$-based MTJ or 160% in its $Fe_3GeTe_2$/h-BN/$Fe_3GeTe_2$-based counterpart is undoubtedly exciting, their operating temperatures of 2 K and 4.2 K, respectively, make such devices impractical. Furthermore, ferromagnetic heterostructures that use either $Fe_3GeTe_2$ or $VSe_2$ suffer from instability, as these materials oxidize quickly upon exposure to air, which could even cause the TMR ratio of such devices to reduce if chemical residues were to subsequently remain on the substrates during the fabrication process [66]. A low TMR ratio reduces the precision with which a neuromorphic system can execute analog operations, just as a small range of channel conductance remains an issue in ferroelectric devices. Both instances can increase the cost of analog-to-digital converters (ADCs) [60]. The above reasons preclude any 2D materials-based MTJ devices from reaching the desired corner in Fig. 3d unless there are breakthroughs in the discovery of novel magnetic 2D materials.

At the algorithm level, the application of 2D materials-based devices for neuromorphic computing has scarcely strayed away from forward-propagation in multi-layer perceptrons (MLPs) using crossbar arrays. NeuroSim+ simulations implement *ex situ* training by performing all gradient computations in software and exporting the updated weights. While NeuroSim+ implements only gradient descent (GD), its newer version MLP+NeuroSimV3.0 offers stochastic gradient descent (SGD), adaptive moment estimation (Adam), and a variant on Polyak's momentum (heavy-ball method), but all of these optimization and relevant acceleration techniques again reside only in software [67]. Therefore, the accuracy results following from NeuroSim+ show only that a device suitably quantizes neural network weights, and the broader crossbar architecture properly executes VMM for forward-propagation. Yet this hardware VMM operation does not include non-linearity computation, as the peripheral circuit for the sigmoid function is not simulated by NeuroSim+. The next generation of neuromorphic systems should depart from inference-based approaches and embrace *in situ* execution of back-propagation techniques, like GD or SGD, thereby paving the way for high-efficiency learning systems that exist purely in hardware.

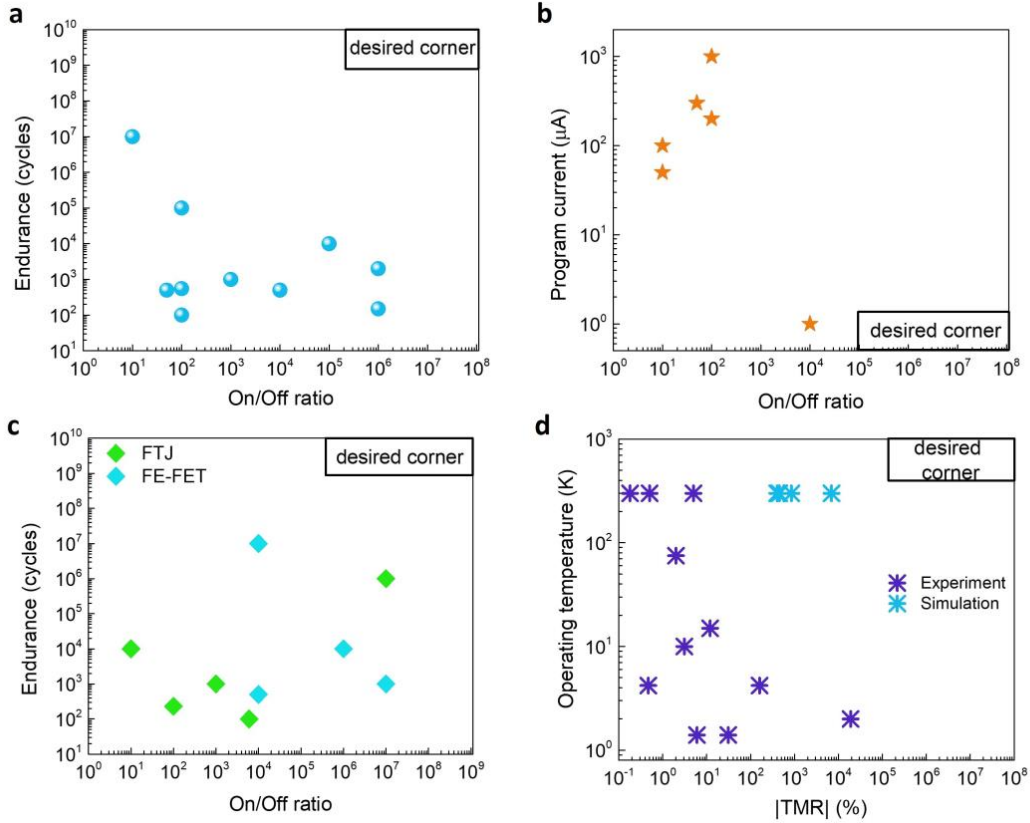

**Fig. 3. Corner plots for 2D materials-based ReRAM, PCM, MRAM, and FeRAM technologies.** (**a**) Cycling endurarance versus ON/OFF ratio of 2D materials-based ReRAM devices [32-36, 68-71]. (**b**) Programming current versus ON/OFF ratio of 2D materials-based PCM devices [37, 39, 40, 72-74]. (**c**) Absolute value of the tunneling magnetorestance (TMR) ratio versus operating temperature of 2D materials-based MRAM devices [52-56, 75-79]. (**d**) Cycling endurance versus ON/OFF ratio of 2D materials-based FeRAM devices [43, 44, 47, 48, 80-84].

## 5. Possibilities

The key benefit of 2D materials in NVM and neuromorphic devices lies in just one major aspect of their physical property: atomically-small thickness. This property leads to two performance advantages in most devices: voltage scaling and gate-tunability. Voltage scaling, while very important and necessary, will inherently compromise endurance and reliability for ReRAM devices. Therefore, 2D materials-based ReRAM devices are unlikely to surpass oxide-based ReRAM for the foreseeable future. However, gate-tunable memristors and memtransistors from 2D materials may provide a unique functional advantage though their device scaling and integration into dense arrays remains undemonstrated [85]. A similar argument could be made for PCM devices based on 2D materials. The space for new development in PCM is more fertile nonetheless, as the gamut of 2D materials keeps expanding, which means the search for electrically-driven crystalline-to-crystalline phase transition in 2D materials should remain an active area of research in the coming years.

The ferroelectric devices community has invested significantly in scaling down the thickness of ferroelectric films, with HZO achieving ferroelectricity at the 1 nm scale [65], but surface-related effects and enhanced depolarization field ultimately prevent traditional ferroelectrics from scaling any further than this fundamental thickness limit [60]. Instead, a vast array of 2D materials have recently been shown to possess ferroelectric properties, including SnTe, $In_2Se_3$, SnS, and CIPS [86-89], which not only allow ferroelectric memory devices to be more aggressively scaled but also offer simpler integration with substrates due to strong intralayer coupling and weak interlayer interaction [60]. Furthermore, non-ferroelectric 2D materials, like $MoS_2$, graphene, and $WSe_2$, have shown superior performance as the active layer in ferroelectric devices that also enable low-dimensional devices and tend to combine well in with traditional ferroelectric materials in a gate stack. These 2D channels can help alleviate reliability problems faced by FE-FETs due to charge traps at the FE-channel interface that often led to charge injection and information loss [60]. Similarly, the ferroelectric-semiconductor-FET (FeS-FET) comprising an intrinsically ferroelectric 2D semiconductor as its channel material resolves many interface and ferroelectric dielectric depolarization issues encountered in traditional FE-FETs [90]. Therefore, research in ferroelectric semiconductors with desirable band gaps, doping, and mobility will remain an active area of research. Likewise, another promising research space will be for MTJs and other magnetic memory devices from 2D materials, particularly searching for novel 2D materials with chemical stability under ambient temperature and $> 300$ K.

With respect to the architecture level, all of the previously mentioned neuromorphic architectures have stopped at achieving spike potentiation and depression. However, a recent report presents the most comprehensive simulation of 2D materials-based devices for actualization-based neuromorphic computing, illustrating the exciting possibility of low-energy *in situ* training for SNNs [91]. The authors not only construct an FE-FET with a 2D graphene channel and PVDF-TrFE ferroelectric but use this device to construct the synapse component of the remote supervise method (ReSuMe). This method was developed for supervised learning in SNNs by applying the Widrow-Hoff rule. While this demonstration is impressive due to its *in situ* training of an SNN in 2D materials-based FE-FETs, the devices themselves are still fairly primitive. Nonetheless, it raises a tantalizing possibility that if the materials and devices were optimized as discussed above, the neural architectures comprising 2D materials-based NVM devices would certainly be possible and bring about a paradigm shift in hardware design in the entire community.

In summary, 2D materials present an exciting new possibility for NVM devices and neuromorphic computing using them. They indeed exhibit functional performance and system-level integration advantages due to their atomic-level thickness and van der Waals nature. In particular, some 2D materials hold strong attributes for PCM and ferroelectric memory devices that promise to surpass all incumbent competitors. However, the expectation of translating such 2D materials-based devices into any useful system either for standard NVM in von Neumann digital computing or analog neuromorphic computing must be tempered until the desired quality of materials is achieved that can enable reliable performance and low variability in devices at scale.

**Acknowledgments:** D.J. and X.L. acknowledge support from Intel RSA. K.K. acknowledges support from the National Science Foundation (NSF) Graduate Research Fellowship Program (GRFP), Fellow ID: 2022338725.

**Author Contributions:** All authors conceived the structure and contents of the manuscript. K.K. and X.L. did the majority of the writing and literature research under supervision of D.J. All authors read, commented, and edited the manuscript.

**Competing Interests:** The authors declare no competing interests.

# References & Notes